\documentclass[pre,aps,floatfix,twocolumn, superscriptaddress]{revtex4-1}

\usepackage{amsmath}

\usepackage{amsfonts}

\usepackage{amssymb}
\usepackage{color,soul}
\usepackage{graphicx}
\usepackage{empheq}
\usepackage{verbatim}

\usepackage{subfigure}

\allowdisplaybreaks 

\begin{document}


\title{Inference of the electron temperature in ICF implosions from the hard X-ray spectral continuum}

\author{Grigory~Kagan}
\email[E-mail: ]{kagan@lanl.gov}
\affiliation{Los Alamos National Laboratory, Los Alamos, NM 87545}
\author{O.~L.~Landen}
\affiliation{Lawrence Livermore National Laboratory, Livermore, CA 94551}
\author{D.~Svyatskiy}
\affiliation{Los Alamos National Laboratory, Los Alamos, NM 87545}
\author{H.~Sio}
\affiliation{Massachusetts Institute of Technology, Cambridge, MA 02139}
\author{N.~V.~Kabadi}
\affiliation{Massachusetts Institute of Technology, Cambridge, MA 02139}
\author{R.~A.~Simpson}
\affiliation{Massachusetts Institute of Technology, Cambridge, MA 02139}
\author{M.~Gatu~Johnson}
\affiliation{Massachusetts Institute of Technology, Cambridge, MA 02139}
\author{J.~A.~Frenje}
\affiliation{Massachusetts Institute of Technology, Cambridge, MA 02139}
\author{R.~D.~Petrasso}
\affiliation{Massachusetts Institute of Technology, Cambridge, MA 02139}
\author{R.~C.~Shah}
\affiliation{Los Alamos National Laboratory, Los Alamos, NM 87545}
\author{T.~R.~Joshi}
\affiliation{Los Alamos National Laboratory, Los Alamos, NM 87545}
\author{P.~Hakel}
\affiliation{Los Alamos National Laboratory, Los Alamos, NM 87545}
\author{T.~E.~Weber}
\affiliation{Los Alamos National Laboratory, Los Alamos, NM 87545}
\author{H.~G.~Rinderknecht}
\affiliation{Lawrence Livermore National Laboratory, Livermore, CA 94551}
\author{D.~Thorn}
\affiliation{Lawrence Livermore National Laboratory, Livermore, CA 94551}
\author{M.~Schneider}
\affiliation{Lawrence Livermore National Laboratory, Livermore, CA 94551}
\author{D.~Bradley}
\affiliation{Lawrence Livermore National Laboratory, Livermore, CA 94551}
\author{J.~Kilkenny}
\affiliation{Lawrence Livermore National Laboratory, Livermore, CA 94551}

\date{\today}

\begin{abstract}
Using the free-free continuum self-emission spectrum at photon energies above 15 keV is one of the most promising concepts for assessing the electron temperature in ICF experiments. However, these photons are due to suprathermal electrons whose mean-free-path is much larger than thermal, making their distribution deviate from Maxwellian in a finite-size hot-spot. The first study of the free-free X-ray emission from an ICF implosion is conducted with the kinetic modifications to the electron distribution accounted for. These modifications are found to result in qualitatively new features in the hard X-ray spectral continuum. Inference of the electron temperature as if the emitting electrons are Maxwellian is shown to give a lower value than the actual one.
\end{abstract}

\maketitle


\section{Introduction}

Knowledge of the temperature of the deuterium-tritium (DT) hot-spot is crucial for success of inertial confinement fusion (ICF) experiments~\cite{lindl1998inertial,atzeni2004physics}. Accurate and reliable temperature measurements constrain theoretical models of the implosions and help identify shortcomings of presently available simulation tools, such as radiation-hydrodynamics codes~\cite{zimmermann1975numerical, marinak2001three}.

Under thermonuclear conditions the fuel is in the form of a plasma and the ion and electron temperatures, $T_i$ and $T_e$, should generally be distinguished. The ion temperature can be inferred from the spectra of the fusion reaction products~\cite{brysk1973fusion, gatu2016indications}. Information about the electron temperature is, \emph{in principle},  carried by the radiation  from the hot imploded plasma, as this radiation is facilitated by free electrons scattering over ions. Since electrons are much faster it is their distribution only that governs the emission spectrum, which should then provide the basis for the $T_e$ inference.  

In practice, such inference is obscured by opacity effects. The photon emitted in the hot-spot has to travel through the hot-spot itself as well as through the remnants of the shell before being registered by spectrometers surrounding the implosion. With a substantial probability of  a reabsorption or scattering event it is a challenge to retrieve the emitting electron distribution from the measured emission spectrum. The current consensus is therefore that the successful diagnostic should operate with the harder part of the spectrum, i.e., photon energies $\hbar\omega \gtrsim$ 15 keV, for which the imploded capsule is transparent~\cite{yaakobi1991diagnosis, ma2012imaging,jarrott2016hotspot,chen2017krypton}. 

In particular, the newly developed spectrometer at the National Ignition Facility (NIF) will infer the electron temperature from the hard X-ray spectral continuum of $20~\mathrm{keV} \lesssim \hbar\omega \lesssim  30~\mathrm{keV}$, which is established through the free-free Bremsstrahlung by electrons scattering off the fully ionized D and T ions~\cite{thorn2017design}.
At the OMEGA laser facility, an electron temperature diagnostic is also being designed based on the time-resolving streak-camera approach~\cite{sio2016particle}, which will look at multiple energy bands in the range of $10~\mathrm{keV} \lesssim \hbar\omega \lesssim  30~\mathrm{keV}$ in cryogenic and warm ICF implosions.



However, while promising the most unambiguous $T_e$ measurements the hard X-ray diagnostic of the hot-spot poses a non-trivial theoretical problem.  For ignition scale implosions with temperatures of about 2.5-5 keV, a photon with $\hbar\omega \gtrsim$ 15 keV can only be produced by a \emph{suprathermal} electron. Its mean-free-path scales as 
\begin{equation}
\label{eq: supra-mfp}
\lambda_{\varepsilon} = \varepsilon ^2 \lambda_0, 
\end{equation}
where $ \lambda_0$ is the standard, thermal mean-free-path and $\varepsilon \equiv E/T_e$ with $E$ being the  electron energy. For $\varepsilon \gtrsim 3$, $\lambda_{\varepsilon} \gg \lambda_0$ and can be comparable to the hot-spot radius $R_h$, making the suprathermal electron distribution deviate from equilibrium even if the bulk electrons are close to Maxwellian. In other words, the hard X-ray diagnostics probe the likely non-thermal tail of the electron distribution, whereas it is the width of the main, thermal part of this distribution that defines $T_e$. 

We present the first calculation of the X-ray spectrum of the free-free self-emission from an ICF hot-spot accounting for the electron kinetic effects. The electron tail depletion is found to affect substantially the harder part of the spectrum to be used for diagnosing the plasma conditions. It is then shown that by assuming Maxwellian electrons one would obtain an electron temperature, which is lower than the actual one.

The remainder of this paper is organized as follows. In the next section we outline the reduced kinetic framework used for evaluating the modified electron distributions. In section~\ref{sec: emission}, technical part of the spectrum calculation is described and essential results are presented. Finally, in section~\ref{sec: discussion}, we suggest a physically intuitive explanation for the main predictions and discuss their practical implications.

\section{Reduced kinetic description for suprathermal electrons}
\label{sec: kinetics}

The problem of the suprathermal particle distribution under conditions relevant to ICF hot-spots has been well explored for ions~\cite{henderson1974burn, petschek1979influence,molvig2012knudsen,albright2013revised,schmit2013tail,tang2014reduced,mcdevitt2014comparative, davidovits2014fusion,cohen2014one,kagan2015suprathermal}. These studies were motivated by the observation that at the temperatures of about several keV most of the fusion reactions involve suprathermal ions, which belong to the so-called ``Gamow window''~\cite{atzeni2004physics}. Due to their larger mean-free-path the suprathermal ions escape from the hot-spot faster than their population can be replenished through collisional up-scattering of thermal ions. Consequently, the tail of the ion distribution is depleted and the fusion reactivity is lower than the normally assumed, Maxwellian averaged value. The associated reduction of the fusion yield has been argued to contribute to degrading the implosion performance in a number of ICF experiments~\cite{rosenberg2014exploration,rinderknecht2015ion,hoffman2015approximate}.

More recent work has also pointed out that the ion burn temperatures, inferred from the width of the fusion reaction product spectra, should be affected by the ion tail depletion as well as the yield~\cite{kagan2015suprathermal}. It has been demonstrated that applying the standard Brysk formula~\cite{brysk1973fusion} to tail depleted ion distributions makes the inferred ion temperature lower than the actual one. Furthermore, it is predicted that the tail depletion makes the apparent DD temperature lower than DT. This trend is indeed observed persistently at NIF, although several other mechanisms can also be responsible for these observations: in particular, residual fluid motion~\cite{appelbe2011production,murphy2014effect,spears2015three}, ion thermal decoupling~\cite{rinderknecht2015ion}, and neutron scattering and burn weighting due to reactivity dependence on temperature~\cite{gatu2016indications}.

The key simplification in the kinetic analysis of the suprathermal particles results from their much fewer number compared to thermal particles. A given suprathermal particle is much more likely to collide with a thermal particle than with another suprathermal one. We also assume that the thermal particle mean-free-path is small, $\lambda_0 \ll R_h$. The suprathermal particles can then be viewed as the test particles scattering over the thermal, Maxwellian background. Within this \emph{reduced} kinetic framework the suprathermal particle distribution is described by a linear partial differential equation (PDE) as opposed to the non-linear integro-differential equation in the full-kinetic approach.
\begin{figure}[]
\begin{center}
\includegraphics[scale=0.8]{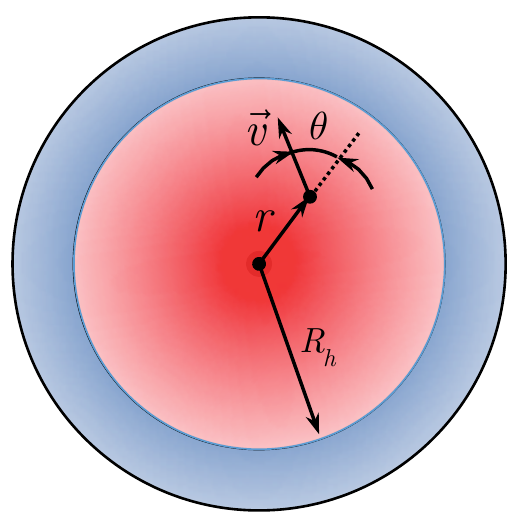}
\caption{ \label{fig: model-sketch}   Sketch of the considered model. Spherically symmetric hot-spot with the radius $R_h$ (in red) is surrounded by cold material (in blue). Due to the symmetry the electron distribution function depends on three variables only: the radial coordinate $r$, the magnitude of the particle velocity $v$ and the angle $\theta$ between the velocity vector $\vec{v}$ and the radial direction.}
\end{center}
\end{figure}

We consider the spherically symmetric problem as sketched in Fig.~\ref{fig: model-sketch}, which also shows the employed variables. With the assumptions outlined in the preceding paragraph the reduced kinetic equation for the suprathermal electrons is structurally identical to that for the suprathermal ions, making it possible to use all the machinery developed in earlier studies~\cite{henderson1974burn, petschek1979influence,molvig2012knudsen,albright2013revised,schmit2013tail,tang2014reduced,mcdevitt2014comparative, davidovits2014fusion,cohen2014one,kagan2015suprathermal}. Here we follow derivation of Ref.~\cite{kagan2015suprathermal} to write
\begin{align}
\nonumber
\mu \frac{\partial f}{\partial x} +  \frac{1- \mu^2}{x}  \frac{\partial f}{\partial \mu} =& \\
\label{eq: kin-eq}
\frac{1}{N_K}  \Bigl[ 
\frac{ 1}{2 \varepsilon^{2}}\frac{\partial }{\partial \mu}  (1&-\mu^2)\frac{\partial f}{\partial \mu}
+
\frac{2 \varkappa }{ \varepsilon} \frac{\partial }{\partial \varepsilon} \Bigl( f + \frac{\partial f}{\partial \varepsilon}          \Bigr)  \Bigr],
\end{align}
where $f$ is the distribution function of the suprathermal electrons, $x \equiv r/R_h$, $\mu \equiv \cos \theta$ and $\varepsilon \equiv E/T_e \equiv m_e v^2/2T_e $ with $m_e$ being the electron mass. The parameter $\varkappa$ is governed by the scattering Maxwellian background of thermal electrons and ions:
\begin{equation}
\label{eq: supra-mfp}
\varkappa =\frac{1}{ 1 + Z_{\bf eff} }, 
\end{equation}
where the effective ion charge number $Z_{\bf eff}$ is defined through
\begin{equation}
\label{eq: eff-charge}
n_e Z_{\bf eff} \equiv  \sum_{\alpha} n_{\alpha} Z_{\alpha}^2
\end{equation}
with $n_{\alpha}$ and $Z_{\alpha}$ being the number density and the charge number, respectively, of the plasma species ``$\alpha$'' and sum is over the ion species only. Generally, $Z_{\bf eff}$ and thus $\varkappa$ depend on the relative concentrations of ion species, but in the practically important case of the DT fuel  $Z_{\bf eff}$ is equal to 1 identically and so $\varkappa_{DT} = 1/2$ for any concentration. 

The parameter $N_K \equiv \lambda_0/R_h$ on the right side of Eq.~(\ref{eq: kin-eq}) is the so-called ``Knudsen number'' which governs the importance of the kinetic effects. Here the thermal electron mean-free-path $\lambda_0$ is defined as $v_{Te}/\nu_e$, where $v_{Te} = \sqrt{2T_e/m_e}$ is the electron thermal speed and
\begin{equation}
\label{eq: coll-freq}
\nu_e =  \frac{4\pi n_e (1+ Z_{\bf eff}) e^4 \ln{\Lambda}}{m_{e}^2 v_{Te}^3}
\end{equation}
is the electron collision frequency with $e$ being the electron charge and $\ln{\Lambda}$ being the Coulomb logarithm.

The kinetic equation~(\ref{eq: kin-eq}) assumes that $T_e$ does not depend on $x$ in the hot-spot. Such an assumption was found to give predictions for main observables, which agree reasonably well with the solutions for realistic temperature profiles in an isobaric hot-spot~\cite{kagan2015suprathermal}. We also notice that  the quasi-stationary description may become invalid for $\lambda_\varepsilon \gg R_h$, since for such large $\varepsilon$ collisions are too rare to have the modified equilibrium established while the bulk plasma parameters can be considered unchanged.

In addition to Eq.~(\ref{eq: kin-eq}) physical distributions should satisfy several constraints. First, they should be isotropic at the center because of the spherical symmetry
\begin{equation}
\label{eq: constraint-sym}
\partial f(x=0,\mu , \varepsilon)/\partial \mu =0.
\end{equation}
Second, there is no suparthermal electron inflow from the cold material into the hot-spot. For the relevant electrons at the interface between the hot and cold plasmas, $0 \leq \theta \leq \pi/2$ and so
\begin{equation}
\label{eq: constraint-current}
f(x=1,-1\le \mu \le 0, \varepsilon) =0.
\end{equation}
Finally, inside the hot-spot $f$ should tend to Maxwellian as $\varepsilon$ approaches $1$ since thermal particles are assumed to be close to equilibrium.

Solution to the PDE problem formulated by Eqs.~(\ref{eq: kin-eq}), (\ref{eq: constraint-sym}) and (\ref{eq: constraint-current}) gives the modified quasi-stationary distribution of suprathermal electrons parametrized by $N_K$. In the next section this solution is used to evaluate the effect of the tail depletion on the X-ray emission spectrum.

\section{Emission spectrum}
\label{sec: emission}

The differential emissivity of a single electron through the free-free Bremsstrahlung process can be written as~\cite{jackson1975electrodynamics, brueckner1976energy,thomas2010abel}
\begin{equation}
\label{eq: emis-single}
\frac{\partial^2 W_s}{\partial\omega \partial t}  \propto \frac{1}{\sqrt{E}} G(E, \omega),
\end{equation}
where $W_s$ is the emitted energy, $\omega$ is the radiation frequency and $t$ is the time. On the right side of Eq.~(\ref{eq: emis-single}), $G(E, \omega)$ is the so-called ``Gaunt factor'', which defines the emissivity as a function of the energies of the emitting electron and emitted photon,  $E$ and $\hbar\omega$, respectively~\cite{gaunt1930continuous}. For the energies of interest, the general quantum-mechanical result for $G(E, \omega)$ can be simplified to become~\cite{greene1959bremsstrahlung}
\begin{equation}
\label{eq: gaunt-born}
G(E, \omega) =
\begin{cases}
    0, & \text{if $E < \hbar\omega$}\\
    \ln \frac{\sqrt{E} + \sqrt{E-\hbar\omega}}{\sqrt{E} - \sqrt{E-\hbar\omega}}, & \text{if $E \geq \hbar\omega$}
  \end{cases}.
\end{equation}

The overall emission spectrum is obtained by integrating the right side of Eq.~(\ref{eq: emis-single}) over the electron distribution and averaging the result over the hot-spot volume. Since the absolute radiation intensity is not important for the present analysis we will be operating with the dimensionless spectrum $S$, which is given by 
\begin{align}
\label{eq: emis-total}
S(\overline{\omega}) =
\frac{(2\pi T_e)^{3/2}}{n_e m_e^{3/2}}
\int \langle f \rangle_{x,\mu} G(\varepsilon, \overline{\omega}) d\varepsilon,
\end{align}
where $ \langle f \rangle_{x,\mu}$ is the distribution function averaged over the $x$ and $\mu$ variables and $\overline{\omega} \equiv \hbar\omega/T_e$ is introduced.
\begin{figure}[h!]
\begin{center}
\includegraphics[scale=0.6]{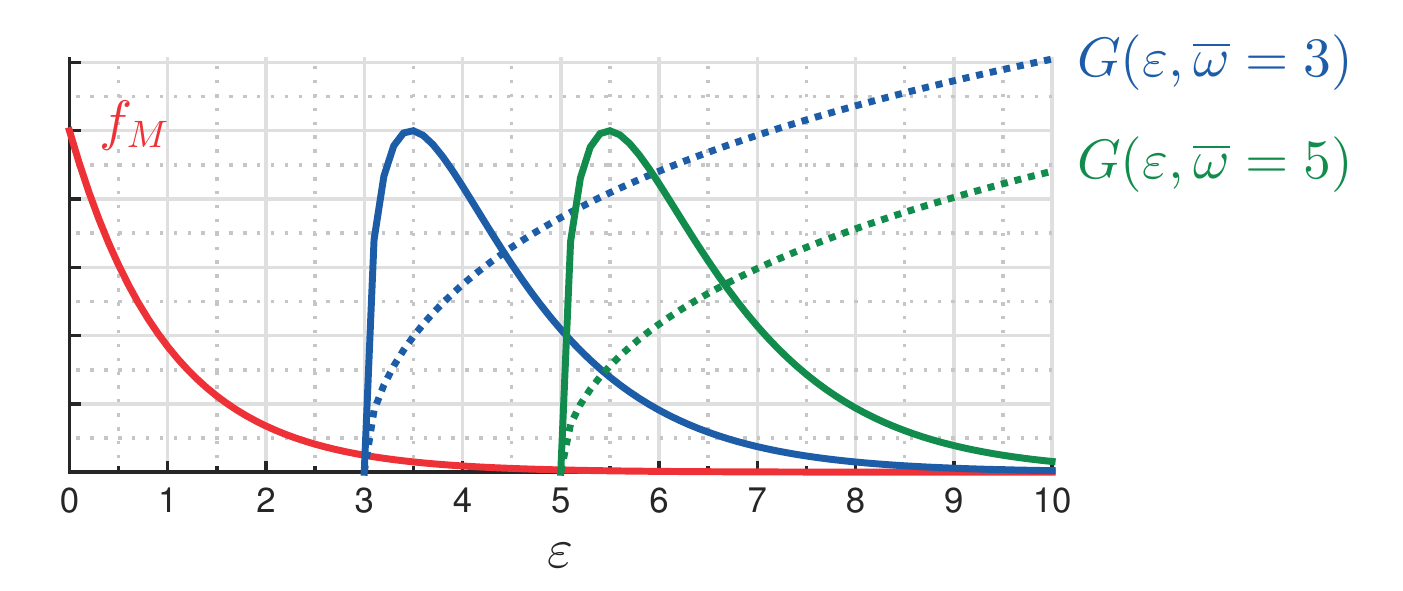}
\caption{ \label{fig: gaunt}   Illustration of the ``Gaunt window'' concept. Red line shows Maxwellian distribution as a function of $\varepsilon$. Dotted blue and green lines show the Gaunt factor~(\ref{eq: gaunt-born}) for $\overline{\omega}$ equal to 3 and 5, respectively, and solid blue and green lines show the corresponding integrand on the right side of Eq.~(\ref{eq: emis-total}) multiplied by appropriate numerical factors for clarity of exposition.}
\end{center}
\end{figure}

To gain an insight into how and when modifications to the distribution function can affect the emission spectrum we first consider Maxwellian distribution 
\begin{align}
\label{eq: Maxwell}
f_M = n_e \Bigl( \frac{m_e}{2\pi T_e} \Bigr)^{3/2} e^{-\varepsilon}.
\end{align}
In Fig.~\ref{fig: gaunt} we plot $f_M(\varepsilon)$ and $G(\varepsilon, \overline{\omega})$ for $\overline{\omega} $ equal to 3 and 5 along with the corresponding integrand from the right side of Eq.~(\ref{eq: emis-total}) (multiplied by appropriate numerical factors to have all the curves exposed). For a given $\overline{\omega}$, the integrand as a function of $\varepsilon$ shows the region of the electron phase space, which contributes the most to the radiation at this frequency. By analogy with the Gamow window, this region can be referred to as the ``Gaunt window''. Since $G(\varepsilon, \overline{\omega})$ is a much slower function of energy than the fusion cross-section $\sigma$, the Gaunt window is more localized and the ``Gaunt peak'' is very close to the lower energy cut-off of $\varepsilon = \overline{\omega}$.

\begin{figure}
\begin{center}
\includegraphics[scale=0.6,trim={0.4cm 1.2cm 1cm 1.3cm},clip]{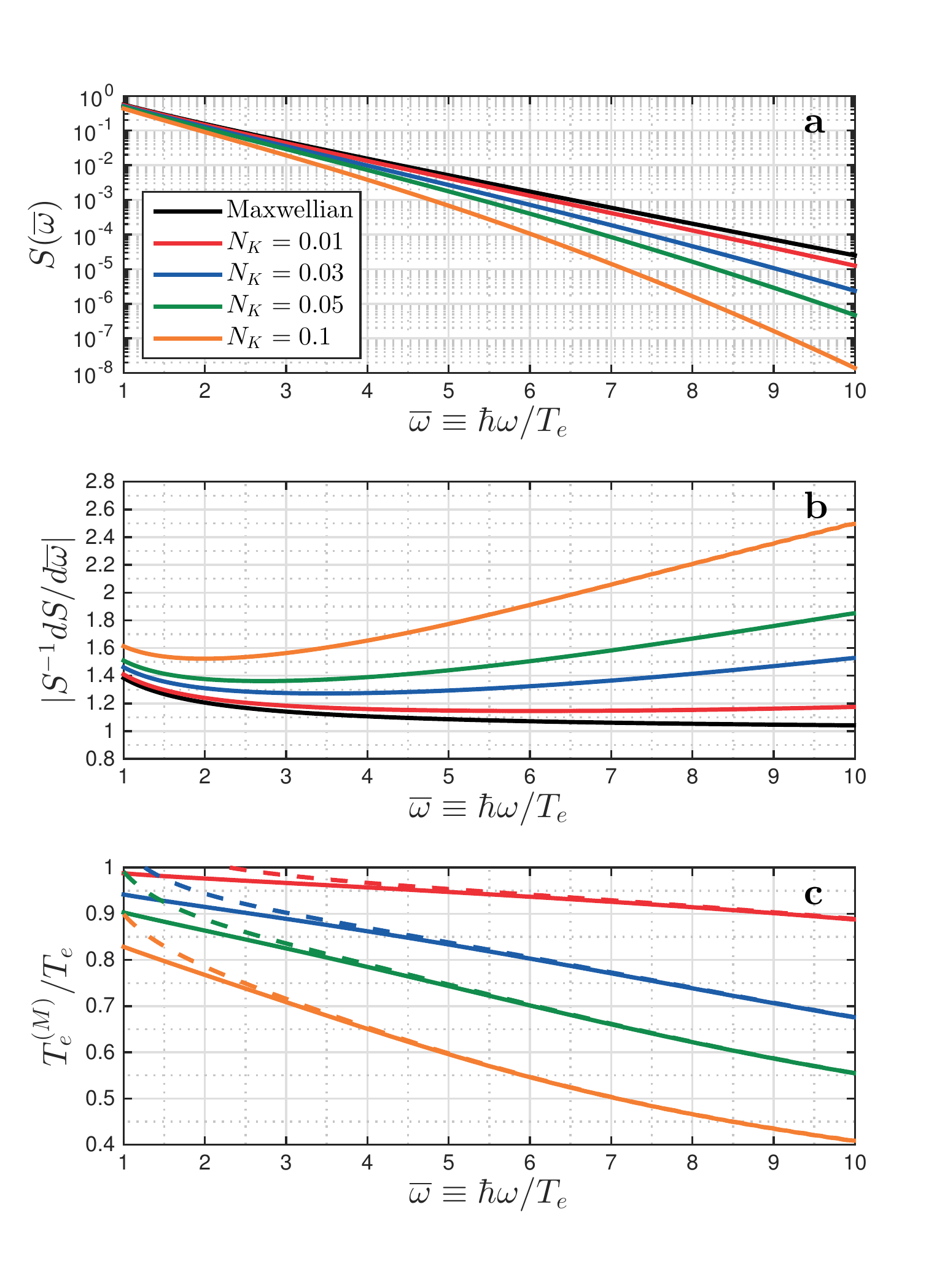}
\caption{ \label{fig: spectrum}   Essential results for Maxwellian and tail depleted electron distributions. {\bf a}: X-ray energy spectrum $S(\overline{\omega})$; {\bf b}:~fractional slope $|S^{-1} dS/d\overline{\omega}|$; {\bf c}: ratio of the inferred and actual temperatures as obtained from the exact, implicit Eq.~(\ref{eq: slope-Maxwell}) (solid line) and approximate, explicit Eq.~(\ref{eq: temp-Maxwell-app}) (dashed line).}
\end{center}
\end{figure}

If within the Gaunt window the electron distribution deviates from equilibrium, the emission spectrum for the corresponding $\overline{\omega}$ should attain non-thermal features. All earlier considered models for the suprathermal particles in the hot-spot have found the tail depletion to be larger for larger energies~\cite{henderson1974burn, petschek1979influence,molvig2012knudsen,albright2013revised,schmit2013tail,tang2014reduced,mcdevitt2014comparative, davidovits2014fusion,cohen2014one,kagan2015suprathermal}. With the electron kinetic effects accounted for one should therefore expect a lower signal for larger $\overline{\omega}$ in the emission spectrum. To quantify this trend we obtain the electron distribution function for several Knudsen numbers $N_K$ as described in the previous section and evaluate the integral on the right side of Eq.~(\ref{eq: emis-total}). The resulting spectra are shown in Fig.~\ref{fig: spectrum}a alongside the computed spectrum for the unperturbed, Maxwellian distribution. The latter can also be calculated analytically to find~\cite{greene1959bremsstrahlung} 
\begin{align}
\label{eq: spectr-Maxwell}
S_M=
e^{-\overline{\omega}/2} K_0(\overline{\omega}/2),
\end{align}
where $K_{j}$ is the $j$th modified Bessel function of the second kind.

We see that, as expected, the spectrum from a tail depleted distribution decays faster than its counterpart from Maxwellian electrons and that the two diverge further faster as $\overline{\omega}$ grows. This feature should have an impact on the apparent electron temperature, since it is the spectrum slope that is supposed to provide $T_e$ in experiments. The slopes are analyzed next with the help of Fig.~\ref{fig: spectrum}b, which shows the logarithmic derivative $|S^{-1} dS/d\overline{\omega}|$ for the spectra from Fig.~\ref{fig: spectrum}a. 

The key finding in Fig.~\ref{fig: spectrum}b is that, compared to the Maxwellian case, the slope dependence on $\overline{\omega}$ for finite Knudsen numbers changes qualitatively. Rather than decreasing monotonically to saturate to 1 with $\overline{\omega}$ becoming large, it reaches a minimum and starts growing. Observing such a trend in experiments can be indicative of a substantial depletion of the electron distribution.

One should then ask the following question of practical importance. If the well resolved spectral continuum is only measured for a relatively narrow range of $\hbar\omega$, it is not possible to see whether the emitting distribution is non-Maxwellian by looking at the experimental analog of Fig.~\ref{fig: spectrum}b. How large of an error in the inferred electron temperature would one make by assuming Maxwellian electrons if their distribution is depleted?

To properly answer this question we first need to specify how $T_e$ is to be inferred when electrons are perfectly Maxwellian. One might expect that for the Maxwellian case the spectrum slope at larger $\hbar\omega$ is mostly due to the exponential decay with $\varepsilon$ of the emitting electron distribution, making the spectrum dependence on $\hbar\omega$ exponential as well. This seems to be supported by the black curve in the semilogarithmic plot in Fig.~\ref{fig: spectrum}a, which appears to be a straight line. 
By fitting a straight line to the experimental data $T_e$ would then be obtained as simply the inverse of the slope with respect to $\hbar\omega$.

However,  according to Fig.~\ref{fig: spectrum}b, such an approach may be not satisfactory. For purely exponential $S(\overline{\omega})$, the logarithmic derivative is constant. While the curve corresponding to Maxwellian in Fig.~\ref{fig: spectrum}b does tend to 1 at large frequencies, it differs from 1 noticeably for practically achievable $\overline{\omega}$. In particular, requiring adequate signal-to-noise ratio is expected to limit spectral measurements to $\overline{\omega} < 6$, where this difference is about 10\% or higher. Since the implosion performance is a very sensitive function of the temperature a special care is thus required when inferring $T_e$ from the spectral data, even if the electrons' deviation from thermodynamic  equilibrium is believed to be negligible.

To retain the non-exponential component of $S(\overline{\omega})$ we recall the exact result~(\ref{eq: spectr-Maxwell}) to find
\begin{align}
\label{eq: slope-Maxwell}
D \equiv
\frac{1}{S_M} \frac{dS_M}{d(\hbar\omega)} =
-\frac{1}{2T_e^{(M)} } \Bigl[1 + \frac{K_1(\hbar \omega/2T_e^{(M)})}{K_0(\hbar \omega/2T_e^{(M)})}  \Bigr].
\end{align}
Using the logarithmic slope $D$, experimentally measured at a certain $\hbar\omega$, for the left side of Eq.~(\ref{eq: slope-Maxwell}) yields a transcendental algebraic equation for $T_e^{(M)}$, \emph{the electron temperature inferred with the assumption of Maxwellian electrons enforced}. One can view Eq.~(\ref{eq: slope-Maxwell}) as implicitly defining $T_e^{(M)}$ as a function of $D$ and $\hbar\omega$.

We can now proceed to evaluating consequences of the tail depletion for the $T_e$ diagnostic. For the depleted distribution the experimentally measured, dimensional slope would be
 \begin{align}
\label{eq: slope-dim}
D = \overline{D} / T_e,
\end{align}
where $\overline{D} \equiv S^{-1} dS/d\overline{\omega}$ is the dimensionless slope whose absolute value is plotted in Fig.~\ref{fig: spectrum}b. We then rewrite Eq.~(\ref{eq: slope-Maxwell}) as 
\begin{align}
\label{eq: inf-temp-1}
\overline{D}  =
-\frac{\hbar \omega/2T_e^{(M)} }{ \hbar \omega/T_e } \Bigl[1 + \frac{K_1(\hbar \omega/2T_e^{(M)})}{K_0(\hbar \omega/2T_e^{(M)})}  \Bigr].
\end{align}
Next we define a function
\begin{align}
\label{eq: h}
h(z)  \equiv
- z \Bigl[1 + \frac{K_1(z)}{K_0(z)}  \Bigr]
\end{align}
and its inverse $H$ such that 
\begin{align}
\label{eq: H}
H(h(z)) \equiv z.
\end{align}
With this in hand, formal solution to Eq.~(\ref{eq: inf-temp-1}) takes the form
\begin{align}
\label{eq: inf-temp-2}
\frac{T_e^{(M)}}{T_e} = 
\frac{\overline{\omega}}{2  H (\overline{D} \overline{\omega} )}.
\end{align}
A point on a given curve in Fig.~\ref{fig: spectrum}b provides the argument $\overline{D} \overline{\omega}$ for the function $H$ on the right side of Eq.~(\ref{eq: inf-temp-2}). This function is then evaluated numerically from the definitions~(\ref{eq: h})-(\ref{eq: H}). Hence, the ratio of the inferred and actual temperatures is obtained from Eq.~(\ref{eq: inf-temp-2}) as shown with solid lines in Fig.~\ref{fig: spectrum}c.

According to this figure, enforcing the assumption of Maxwellian electrons makes the inferred temperature lower than the actual one. The error is larger for higher Knudsen numbers and photon energies, which one might expect: for a given $N_K$ the radiation at larger $\overline{\omega}$ is due to electrons with larger energies and thus larger mean-free-paths, enhancing the kinetic effects. Similarly, for a given emission frequency, the number of emitting electrons becomes smaller as $N_K$ grows. With a more in-depth qualitative analysis one can also suggest an intuitive explanation for the sign of the error. This consideration and further discussion of the results of Fig.~\ref{fig: spectrum}c are presented in the next section. We conclude this section by presenting a simpler, explicit, and yet quite accurate, formula for $T_e^{(M)}$.

To obtain it, we utilize the asymptotic expansion for the Bessel functions on the right side of Eq.~(\ref{eq: slope-Maxwell}):
\begin{align}
\label{eq: Bessel-Maxwell}
\frac{K_1(\hbar \omega/2T_e^{(M)})}{K_0(\hbar \omega/2T_e^{(M)})} =
1+ \frac{T_e^{(M)}}{\hbar \omega}  + ...
\end{align}
Solving Eq.~(\ref{eq: slope-Maxwell}) with respect to $T_e^{(M)}$ then gives
\begin{align}
\label{eq: temp-Maxwell-app}
T_e^{(M)} \approx \frac{2\hbar\omega}{2\hbar\omega |D| - 1},
\end{align}
where we used that the slope is negative, $D = - |D|$. Within the same approximation Eq.~(\ref{eq: inf-temp-2}) becomes 
\begin{align}
\label{eq: inf-temp-3}
\frac{T_e^{(M)}}{T_e} \approx 
\frac{2 \overline{\omega}}{2  \overline{\omega}  |\overline{D}| - 1}.
\end{align}
We have applied this formula to the slope data of Fig.~\ref{fig: spectrum}b and plotted the results with dashed lines in Fig.~\ref{fig: spectrum}c. It can be observed that the explicit expression~(\ref{eq: temp-Maxwell-app}) does provide an  accurate approximation for the exact, implicit relation~(\ref{eq: slope-Maxwell}) for $ \overline{\omega} \gtrsim 3 $.

\section{Discussion}
\label{sec: discussion}


\begin{figure}[h!]
\begin{center}
\includegraphics[scale=0.6,trim={7.1cm 4.6cm 6.7cm 4.6cm},clip]{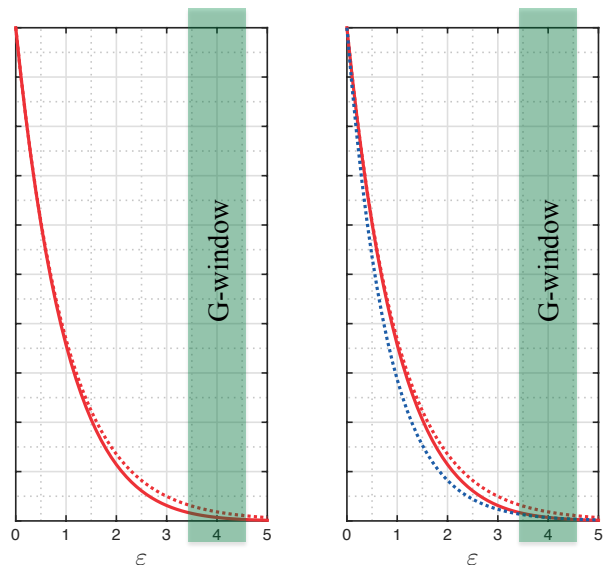}
\caption{ \label{fig: interp}  Sketch for the qualitative explanation of the difference between the actual temperature and the temperature inferred with the assumption of Maxwellian distribution enforced. Solid and dotted red lines denote, respectively, the actual, depleted distribution and Maxwellian distribution with the bulk temperature. The semi-transparent green rectangle denotes the Gamow window (in the case of the nuclear diagnostics) or Gaunt window (in the case of the hard X-ray diagnostics). On the right, the dotted blue line shows the Maxwellian reconstructed from the piece of the depleted distribution inside the G-window.}
\end{center}
\end{figure}

Essential quantitative predictions for the emission spectrum are presented in Fig.~\ref{fig: spectrum} and depend on two key parameters, the photon energy with respect to the electron temperature $\overline{\omega} \equiv \hbar\omega/T_e$ and the Knudsen number $N_K \equiv \lambda_0/R_h$. 
To gauge the importance of the discussed physics for ICF experiments one needs an estimate for them.

In exploding pusher implosions, the Knudsen number is routinely found to be a fraction of unity or even larger with $T_e$ in the 5 to 10 keV range~\cite{rosenberg2014exploration}. The case of $N_K = 10\%$, which is shown with the orange line in Fig.~\ref{fig: spectrum}, should therefore provide a conservative estimate. To remain on the conservative side we also pick $T_e \approx 10$~keV. It can then be seen from Fig.~\ref{fig: spectrum}c that the electron temperature inferred from the spectral slope at 20 keV ($\overline{\omega} \approx 2$) would be about $20\%$ lower than the actual one if modifications to the electron distribution function are not accounted for. Higher Knudsen numbers, lower electron temperatures and measuring the slope at higher photon energies would all make the error grow further.

In ignition scale implosions at NIF, $N_K \sim 1\%$. With $T_e$ about 5 keV, the corresponding curve in Fig.~\ref{fig: spectrum}c predicts an error of 4 to 7\% in the inferred temperature for the slopes measured at 20 to 30 keV. However, practical ignition scale implosions at NIF deviate strongly from the simple 1D design, which will likely enhance the effect substantially. 

That is, the Knudsen number defined as $\lambda_0/R_h$ for the kinetic analysis of Sec.~\ref{sec: kinetics} becomes irrelevant for 2D and 3D implosions. Solving the reduced kinetic equation in such cases is greatly complicated.  Yet, one can use the 1D results of Fig.~\ref{fig: spectrum} to estimate the inferred temperature in realistic geometries by defining the effective Knudsen number $N_K^{(\bf eff)} \equiv  \langle \lambda_0/L  \rangle$, where $L$ is the characteristic spatial scale and $ \langle ...  \rangle$ denotes the volume average. In 2D and 3D implosions, scales much smaller than the hot-spot radius $R_h$ appear and, equally importantly, occupy the outer radii that constitute a large fraction of the hot-spot volume. One can thus expect that $N_K^{(\bf eff)}$ is at least a few times larger than the ``nominal'' Knudsen number $\lambda_0/R_h$. In turn, according to Fig.~\ref{fig: spectrum}c, even for the Knudsen number of $3\%$ the effect on the temperature becomes very noticeable.

It should also be noted that the predicted effect of the electron tail depletion on the temperature inferred from hard X-ray spectral continuum is in the same direction as the earlier considered effect of the ion tail depletion on the temperature inferred from the spectrum of the fusion reaction products~\cite{kagan2015suprathermal}. We suggest a physically intuitive explanation for this similarity with the help of Fig.~\ref{fig: interp}.

There, we first sketch on the left plot with solid and dotted red lines, respectively, the actual, depleted distribution and the Maxwellian distribution corresponding to the bulk particle temperature $T$. The difference between the two only becomes substantial for particles several times more energetic than $T$. Since such particles constitute only a small fraction by number the temperature moment for the depleted distribution is close to $T$.

However, the nuclear and hard X-ray diagnostics assign a disproportionally large statistical weight to suprathermal particles which belong to the Gamow and Gaunt windows, respectively, where the difference between the two distributions is more pronounced. In  Fig.~\ref{fig: interp} such a window is sketched with the semi-transparent green rectangle. Only the part of the solid red curve covered by this window contributes to measurements. Assumption of Maxwellian particles effectively means that the diagnostics reconstruct the Maxwellian distribution from the piece inside the window back into the region of smaller $\varepsilon$, as it is shown by the dotted blue line in the right plot in Fig.~\ref{fig: interp}. Since the dotted blue curve is lower than the dotted red curve for larger $\varepsilon$, it gives a more narrow distribution in the bulk region of the phase space which defines the temperature. Consequently, the inferred temperature turns out to be smaller than the actual one.

Finally, we notice that the similarity of the mechanisms behind the suprathermal ion and electron distributions can make the  spectral measurements very instrumental for studying kinetic effects in ICF implosions. So far, experimental observations of these effects have mostly been through the fusion yields. For a given fusion reaction, the yield carries cumulative information about a certain region of the particle phase space, namely, the Gamow window. To probe other regions---i.e., to ``shift'' the Gamow window---one can only look at a different fusion reaction. The number of different reactions in an ICF hot-spot is very limited, however, and  detailed studies of the particle distribution with the nuclear diagnostics are hardly possible. On the other hand, the Gaunt window can be shifted in a controlled fashion by looking at different emission frequencies. Hence, well resolved measurements of the free-free spectral continuum has the potential of providing unprecedented information about kinetic processes in the hot DT core.

\acknowledgements
The work was supported by the Laboratory Directed Research and Development program under the auspices of the U.S. Dept. of Energy by the Los Alamos National Security, LLC, Los Alamos National Laboratory under Contract No. DE-AC52-06NA25396.


\end{document}